\documentclass[a4paper]{jpconf}
\usepackage{graphicx}
\begin{document}
\title{On spectrum of  vacuum energy}

\author{G.E.~Volovik}

\address{ Low Temperature Laboratory, Helsinki University of
Technology, Finland\\
L.D. Landau Institute for Theoretical Physics,
 Moscow, Russia}

\ead{volovik@boojum.hut.fi}

\begin{abstract}
We discuss the problem of the spectral function of vacuum energy. In traditional approach the ultraviolet divergencies of the vacuum energy are cancelled by imposing relations between different quantum fields and their masses. The emergent theories suggest that the microscopic degrees of the underlying quantum vacuum add to the spectral function and their contribution  cancels the diverging zero point energy of quantum fields. Examples of the spectral function
of the vacuum energy in the condensed-matter systems with relativity emerging at low energy 
are presented. In the Sakharov induced gravity situation may be even more dramatic: only microscopic (Planck scale) constituent fields contribute to the vacuum energy, while the  diverging zero-point energy of emergent quantum field (gravitational field) is missing. On the other hand consideration of  the 
fermionic condensed matter systems suggests that emergent relativistic fermionic quasiparticles
contribute in conventional way as Dirac vacuum. 

\end{abstract}

\section{Introduction} 

The thermodynamic approach~
\cite{KlinkhamerVolovik2008a,KlinkhamerVolovik2008b,KlinkhamerVolovik2008c} 
to the cosmological constant problems assumes that
the quantum vacuum  is a stable self-sustained equilibrium state,
which is described by compressibility and
other characteristics of the response to external perturbations.
In this approach, the vacuum energy
density appears in two forms.
First, there is the microscopic vacuum energy density $\epsilon\sim E_{\rm UV}^4$
characterized by an ultraviolet energy scale $E_{\rm UV}$.
Second, there is the macroscopic vacuum energy density which is
determined by a particular thermodynamic potential,
 $\tilde{\epsilon}_{\rm vac}(q) \equiv \epsilon - q\,d\epsilon/dq$,
 where $q$ is a microscopic variable describing the physics of the
ultraviolet vacuum. Thermodynamics and dynamics of $q$ are
described by macroscopic equations, because $q$ is a conserved quantity
similar to the particle density in liquids, which
describes a microscopic quantity -- the density of atoms -- but obeys
the macroscopic equations of hydrodynamics, because of particle-number
conservation.
Different from known
liquids, the quantum vacuum is Lorentz invariant, and the
quantity $q$ is Lorentz invariant.
For a self-sustained vacuum in full thermodynamic equilibrium and
in the absence of matter, the vacuum energy
density $\tilde{\epsilon}_{\rm vac}$ is automatically nullified
(without fine tuning) by the spontaneous
adjustment of the vacuum variable $q$ to its equilibrium value $q_0$,
so that $\tilde{\epsilon}_{\rm vac}(q_0)=0$.
It is important that $\tilde{\epsilon}_{\rm vac}$
contributes to
the effective gravitational field equations as cosmological constant $\Lambda$.
This implies that the effective cosmological constant $\Lambda$
of a perfect quantum vacuum is strictly zero.
 This result follows from the stability of the quantum vacuum in the underlying microscopic theory, but it does not depend on details microscopic structure of the vacuum  from which gravity emerges.

\begin{figure}
\includegraphics[width=1.0\textwidth]{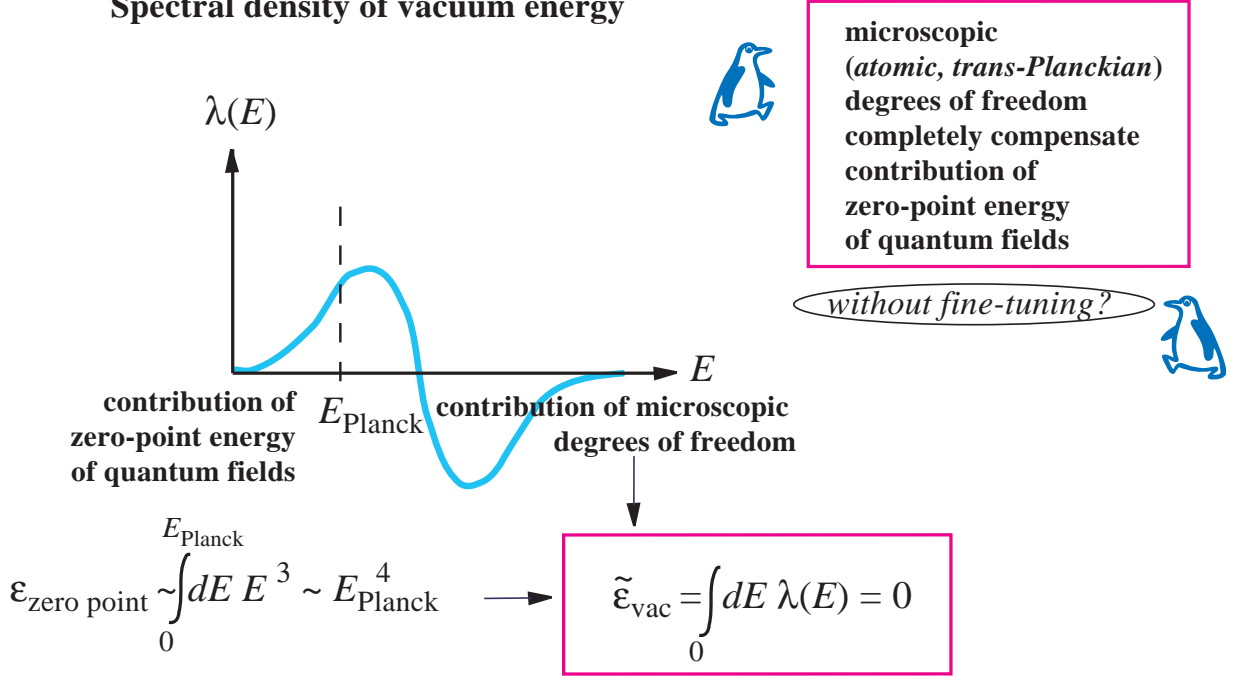}
\caption{Sectral density of $\Lambda$ suggested for the vacuum 
in which bosonic quantum fields  dominate in the sub-Planckain region (from the talk in Ref. \cite{VolovikTalk2005}). Zero point energy of bosonic fields gives rise to the  contribution 
to the energy  of quantum vacuum
which diverges  as $E^4$ and is estimated as $E_{\rm P}^4$, where 
$E_{\rm P}$ is Planck energy scale. In the equilibrium self-sustained quantum vacuum
 \cite{KlinkhamerVolovik2008a,KlinkhamerVolovik2008b,KlinkhamerVolovik2008c} 
this contribution is fully compensated by microscopic (trans-Planckian) degrees of freedom, so that the resulting $\Lambda=\int dE ~\lambda(E)=0$.}
\label{LambdaSpectrum}
\end{figure}

The problem which we discuss here is the fine structure of the vacuum: how different degrees of freedom, macroscopic (sub-Planckian) and microscopic (super-Planckian), contribute to the vacuum energy. For that one may  introduce  the spectral density $ \lambda(E)$ which describes contributions of different energy scales 
$E$ to the vacuum energy:
\begin{equation}
\Lambda = \int_0^\infty  dE ~\lambda(E)~,
\label{SpectralDensity}
\end{equation}
Fig. \ref{LambdaSpectrum} shows the spectral density $ \lambda(E)$ suggested in Ref. \cite{VolovikTalk2005}. 

According to this suggestion, at low energy the dominating contribution to the vacuum energy comes from zero point energy of macroscopic bosonic quantum fields and/or from the occupied negative energy levels of fermionic fields. This contribution quartically diverges  and approaching the Planck scale produces the huge vacuum energy:
\begin{equation}
\ \int_0^{\rm P} dE ~\lambda(E)\sim \int_0^{\rm P}  dE~E^3\sim E_{\rm P}^4~.
\label{ZPE}
\end{equation}
However, at higher energy scale $E\sim E_{\rm P}$, the microscopic degrees of the quantum vacuum intervene and finally compensate the contribution of quantum  fields. As a result, the spectral density satisfies the sum rule  
\begin{equation}
\int_0^\infty  dE~ \lambda(E)=0~.
\label{SpectralDensitySumRule}
\end{equation}
Such a ÒsymmetryÓ between sub- and super-Planckian degrees of freedom \cite{Nobbenhuis2006} is generated by thermodynamics of the equilibrium self-sustained vacuum.
  
In  Refs.  \cite{Kamenshchik2007,Alberghi2008} the other the sum rules were introduced,
which impose the relations between masses of fermionic and bosonic particles in the Standard Model.
We discuss these sum rules using our experience with condensed matter systems, and also the Frolov-Fursaem scheme of Sakharov induced gravity
\cite{FrolovFursaev1998,FFZ2003} which uses the relation between the masses of constituent fields.
  
  \section{Contribution of macroscopic quantum fields to $\Lambda$} 

In Refs.  \cite{Kamenshchik2007,Alberghi2008} the spectral density of the relativistic vacuum has been introduced in a different form which takes into account zero point energy of different quantum fields:
\begin{equation}
 \Lambda=\int \frac{d^3p}{(2\pi)^3}  \int_0^\infty  dx~\rho(x)\sqrt{x+p^2}~.
\label{Spectral}
\end{equation}
Here $\rho(x)$ is a kind of density of states of modes with mass $M=x^{1/2}$. The same spectral function was introduced earlier by Zeldovich \cite{Zeldovich1968}. The contribution of a bosonic scalar field with mass $M_b$ to the spectral function  is $\rho_b(x)=(1/2)\delta(x-M_b^2)$, while the contribution of one spin component of the fermionic Dirac field with mass $M_f$ is $\rho_f(x)=-\delta(x-M_f^2)$. Summation over all bosonic and fermionic fields gives the spectral function
\begin{equation}
\rho_{\rm macro}(x)= \frac{1}{2}\sum_b\delta(x-M_b^2) -  \sum_f\delta(x-M_f^2)  ~.
\label{SpectralFunctionMacro}
\end{equation}  
In this representation the spectral function $\lambda(E)$ in Eq.(\ref{SpectralDensity}) is expressed via the spectral density $\rho(x)$ as:
\begin{equation}
\lambda(E)  =\frac{E^2}{\pi^2}  \int_0^E dp ~p^2 \rho(E^2-p^2)~.
\label{SpectralDensityE}
\end{equation}
In the energy range $M_{b,f}\ll E\ll E_{\rm UV}$ one has the conventional zero-point contribution to the vacuum energy:
\begin{equation}
\lambda_{\rm macro}(E)= \left( \frac{1}{2}N_b-N_F \right) \frac{E^3}{2\pi^2} ~~,~~M\ll E\ll E_{\rm UV}~,
\label{Macro}  
\end{equation}
where $N_b$ and $N_F$ denote the number of bosonic and fermionic species (which properly includes summation over spin components).
 In this representation, the nullification of $\Lambda$ imposes several constraints (sum rules)  for $\rho(x)$:
\begin{equation}
\int_0^\infty dx \rho(x)=  \int_0^\infty dx ~x\rho(x)=\int_0^\infty dx ~x^2\rho(x)=\int_0^\infty dx~x^2\ln x ~\rho(x)=0~.
\label{SumRules}
\end{equation}  
While there is only one condition for $\lambda(E)$ in Eq.(\ref{SpectralDensitySumRule}), there are four conditions for $\rho(x)$. The first three conditions in  (\ref{SumRules}) nullify the quartically, quadratically and logarithmically divergent contributions to the vacuum energy respectively  \cite{Zeldovich1968,Kamenshchik2007}, while the last condition nullifies the rest finite contribution, which only depends on masses of fields.

These sum rules (\ref{SpectralFunction}) give the following relation between bosonic and fermionic species and their masses:
 \begin{eqnarray}
 \frac{1}{2}N_b=N_F~,
\label{Rule1}
\\
\sum_b M_b^2=2\sum_f M_f^2~,
\label{Rule2}
\\
\sum_b M_b^4=2\sum_f M_f^4~,
\label{Rule3}
\\
\sum_b M_b^4\ln M_b=2\sum_f M_f^4 \ln M_f~.
\label{Rule4}
\end{eqnarray}
As before, the first three conditions (\ref{Rule1}-\ref{Rule3}) remove the diverging terms in vacuum energy  \cite{Kamenshchik2007}, while the last condition (\ref{Rule4}) nullifies the cosmological constant.
It was suggested  \cite{Kamenshchik2007,Alberghi2008}
that these sum rules impose relations between the masses of the 
 fermionic and bosonic fields in Standard Model. This is probably valid if 
 all these fields are fundamenta. However this suggestion is not supported in emergent theories.

  \section{Contribution of microscopic degrees of deep vacuum} 
  \label{Contribution_deep_vacuum} 
  
In the emergent theory  \cite{Volovik2003,VolovikTalk2005}, the  bosonic and fermionic  fields are considered as effective low-energy fields, and in addition there are microscopic degrees of freedom of the deep vacuum which should also contribute to the spectral function of the vacuum energy:
\begin{equation}
\rho(x)= \frac{1}{2}\sum_b\delta(x-M_b^2) -  \sum_f\delta(x-M_f^2) +\rho_{\rm micro}(x) ~.
\label{SpectralFunction}
\end{equation}  
The microscopic  contribution $\rho_{\rm micro}(x) $ is determined by  physics of extreme ultraviolet
and thus involves the ultraviolet scale $E_{\rm UV}$, which in principle can be larger that the 
Planck scale. Let us consider for example $\rho_{\rm micro}(x)$  as continuous function of $x$.  Then the dimensional analysis gives the following estimates:  $\rho_{\rm micro}(0)\sim  \rho_{\rm micro}(E_{\rm UV}^2)\sim  E_{\rm UV}^{-2}$.

At low energy $E\ll E_{\rm UV}$, the contribution of microscopic degrees to the spectral function is:
\begin{equation}
 \lambda_{\rm micro}(E)\approx \frac{E^5}{3\pi^2} \rho_{\rm micro}(0) \sim \frac{E^5}{E_{\rm UV}^2}~~,~~E\ll E_{\rm UV}~.
\label{Micro}  
\end{equation}
 This contribution is small compared to the contribution of zero point energy of macroscopic quantum fields in Eq.(\ref{Macro}).
But  at the microscopic energy scale the contribution of the microscopic physics becomes comparable with the macroscopic contribution and  finally it should compensate the zero point energy nullifying $\Lambda$ as required by thermodynamics. The sum rules in Eq.(\ref{SumRules}) become
 \begin{eqnarray}
  \frac{1}{2}N_b-N_F +\int_0^\infty dx \rho_{\rm micro}(x)=0 ~,
\label{NRule1}
\\
 \frac{1}{2}\sum_b M_b^2- \sum_f M_f^2 +  \int_0^\infty dx ~x\rho_{\rm micro}(x)=0~,
\label{NRule2}
\\
 \frac{1}{2}\sum_b M_b^4- \sum_f M_f^4 +\int_0^\infty dx ~x^2\rho_{\rm micro}(x)=0~,
\label{NRule3}
\\
 \frac{1}{2}\sum_b M_b^4\ln M_b-\sum_f M_f^4 \ln M_f +\int_0^\infty dx~x^2\ln x ~\rho_{\rm micro}(x)=0~.
 \label{NRule4}
\end{eqnarray}
Contrary to conditions (\ref{Rule1}-\ref{Rule4}), they do not require any relations between the masses
for masses of fields. 

Let us take into account that the magnitude of  $\rho_{\rm micro}$  is $|\rho_{\rm micro}| \sim E_{\rm UV}^{-2}$ and
 the characteristic scale of the argument  of $\rho_{\rm micro}(x)$ is $x\sim E_{\rm UV}^2$. Then since $(M_b,M_f) \ll E_{\rm UV}$, the masses of fields can be neglected in Eqs.(\ref{NRule2}-\ref{NRule4}) and in the main approximation (i.e. in the limit of zero masses) the conditions become:
 \begin{eqnarray}
  \frac{1}{2}N_b-N_F  +\int_0^\infty dx \rho_{\rm micro}^{(0)}(x)=0 ~,
\label{N0Rule1}
\\
\int_0^\infty dx ~x\rho_{\rm micro}^{(0)}(x)=\int_0^\infty dx ~x^2\rho_{\rm micro}^{(0)}(x)=\int_0^\infty dx~x^2\ln x ~\rho_{\rm micro}^{(0)}(x)=0~.
\label{N0Rule234}
\end{eqnarray}
Correction to  the microscopic function $\rho_{\rm micro}(x)$ due to masses of quantum fields can be estimated comparing Eq.(\ref{NRule2}) with Eq.(\ref{N0Rule234}):
\begin{equation}
\rho_{\rm micro} =\rho_{\rm micro}^{(0)}+ \delta \rho_{\rm micro} ~~,~~ \frac{\delta \rho_{\rm micro}}{\rho_{\rm micro}^{(0)}}  \sim   \frac{M^2}{E_{\rm UV}^2}\ll 1~.
 \label{CorrectionA}
\end{equation}  
This estimate is consistent with the variation of another microscopic parameter -- the Newton constant, $\delta G/G\sim M^2/E_{\rm UV}^2$  (see discussion in Ref. \cite{KlinkhamerVolovik2008a}).

 \section{Examples of microscopic contributions} 

Let us consider several  examples of the microscopic contribution $\rho_{\rm micro}(x)$ to the spectral function. The first two examples represent some simple choices of the continuous function $\rho_{\rm micro}(x)$, which explicitly demonstrate the magnitudes of parameters involved. The third one is the induced gravity discussed by Frolov and Fursaev \cite{FrolovFursaev1998}.

Let us choose $\rho_{\rm micro}(x)$ as a sum of exponents:
\begin{equation}
\rho_{\rm micro}(x)=  \frac{1}{E_{\rm UV}^2}\sum_n a_n\gamma_n\exp(-\gamma_n \frac{x}{E_{\rm UV}^2})~,
\label{Exponents}
\end{equation} 
with parameters $|a_n|\sim  \gamma_n \sim 1$. 
The factor $\rho_{\rm micro}(0)$ which enters the low-energy part of the microscopic contribution in Eq.(\ref{Micro}) is   $\rho_{\rm micro}(0)=E_{\rm UV}^{-2}\sum_n a_n\gamma_n$.

Sum rules in Eqs.(\ref{NRule1}-\ref{NRule4}) are
 \begin{eqnarray}
\frac{1}{2}N_b-N_F +\sum_n a_n=0 ~,
\label{ENRule1}
\\
 \frac{1}{2}\sum_b M_b^2- \sum_f M_f^2 +E_{\rm UV}^2 \sum_n\frac{a_n}{\gamma_n}=0~,
\label{ENRule2}
\\
 \frac{1}{2}\sum_b M_b^4- \sum_f M_f^4 +2E_{\rm UV}^4 \sum_n\frac{a_n}{\gamma_n^2}=0~,
\label{ENRule3}
\\
 \frac{1}{2}\sum_b M_b^4\ln M_b-\sum_f M_f^4 \ln M_f +2E_{\rm UV}^4\sum_n\frac{a_n}{\gamma_n^2}\left(\frac{3}{2}- C +\ln \frac{1}{\gamma_n}\right)=0~,
\label{ENRule4}
\end{eqnarray}
where $C$ is Euler's constant.
In the main approximation $(M_b^2,M_f^2) \ll  E_{\rm UV}^2$  one has
 \begin{eqnarray}
  \frac{1}{2}N_b-N_F +\sum_n a_n^{(0)}=0 ~,
\label{EN0Rule1}
\\
 \sum_n\frac{a_n^{(0)}}{\gamma_n^{(0)}}=  \sum_n\frac{a_n^{(0)}}{\left(\gamma_n^{(0)}\right)^2}=\sum_n\frac{a_n^{(0)}}{\left(\gamma_n^{(0)}\right)^2} \ln \frac{1}{\gamma_n^{(0)}}=0~.
\label{EN0Rule234}
\end{eqnarray}
Solutions of Eqs.(\ref{EN0Rule1},\ref{EN0Rule234}) exist only for the number of exponents $n>3$.

The corrections to the parameters $a_n$ and $\gamma_n$ due to masses  are consistent with
Eq.(\ref{CorrectionA}): from Eq.(\ref{ENRule2}) it follows that $\delta a_n/a_n^{(0)} \sim \delta \gamma_n/ \gamma_n^{(0)}\sim M^2/E_{\rm UV}^2$.
The same properties are shared by another simple choice for $\rho_{\rm micro}(x)$  as a polynomial with argument restricted by the microscopic energy scale:
\begin{equation}
\rho_{\rm micro}(x)= \frac{N_F -\frac{1}{2}N_b}{E_{\rm UV}^2}\theta(E_{\rm UV}^2-x)
\sum_{n=0}^3a_n\frac{x^n}{E_{\rm UV}^{2n}} ~.
\label{Polinom}
\end{equation} 

 \section{Induced gravity: absence of zero-point energy for graviton?} 

The third example is the Frolov-Fursaev scheme of Sakharov induced gravity \cite{FrolovFursaev1998,FFZ2003,Visser2002}. In this theory, the microscopic degrees of freedom  are represented by constituent fermionic and bosonic quantum fields, whose masses  are of order of Planck energy scale. The macroscopic physics contains only one effective field -- gravity. The action for the induced gravitational field is obtained by integration over the constituent quantum fields. 
Masses of constituent fields satisfy 6 equations.  Four of them,  Eqs.(2.16) in Ref. \cite{FrolovFursaev1998}, nullify divergent and finite contributions to the cosmological constant, and the other two in Eq.(2.17) in Ref. \cite{FrolovFursaev1998} are needed for nullification of the ultraviolet divergences in the induced Newton constant $G$. The first four equations coincide with conditions (\ref{Rule1}-\ref{Rule4}). This would indicate that  in the Frolov-Fursaev scheme the spectral function for the vacuum energy is
\begin{equation}
\rho(x)=\rho_{\rm micro}(x)= \frac{1}{2}\sum_b\delta(x-M_{cb}^2) -  \sum_f\delta(x-M_{cf}^2)  ~,
\label{SpectralFunctionFF}
\end{equation}  
where $M_{cb}$  and $M_{cf}$ are masses of constituent bosons and  constituent fermions respectively. Note that the other two conditions  in Eq.(2.17) of Ref. \cite{FrolovFursaev1998}, which are determined by Seeley-DeWitt coefficients, discriminate between vector and scalar bosonic fields. The resulting  six conditions require existence of the all three types of constituent fields in the quantum vacuum: vector, spinor and scalar fields (see Eq.(2.18) in Ref. \cite{FrolovFursaev1998}). But this is not important when only the vacuum energy is considered.

What is important for us is that  all the masses of microscopic constituent  fields are of Planck scale,   $M_{cb}\sim M_{cf} \sim E_{\rm P}$. As a result, the spectral function $\rho(x)$ is identically zero in the sub-Planckian region of energies: $\rho(x\ll E_{\rm P}^2)\equiv 0$. 
This would mean that at low energy the spectral function  $\lambda(E)$ is also  identically zero: $\lambda(E \ll E_{\rm P})\equiv 0$. At first glance, this is a rather surprising result: there is no contribution $\rho_{\rm macro}(x)$ of the low-energy gravitational field to the vacuum energy, i.e.  zero point energy of gravitons is missing contrary to expectation that the contribution of graviton should be the same as that of  two massless scalar fields  corresposponding to two values of the helicity. On the other hand,  in induced theory,  graviton is a composite object made of microscopic fields. It would be strange if, say, hydrogen atom composed of `fundamental'  particles -- quarks and leptons -- will produce diverging terms to the vacuum energy: such a composite object exists only below the scale of compositeness (atomic scale), and thus cannot contribute to the vacuum energy at the scale of the ultraviolet cut-off.

 What should one expect from composite objects in a more realistic approach?  There are several possibilities. One of them is that  zero point energy of graviton does not contribute to the vacuum energy (which does not mean, however, that the Casimir effect is absent for the gravitational waves, since we considered the energy of a homogeneous vacuum). In this case, the low-energy part of $\lambda(E)$ in Eq.(\ref{Micro}) starts with  $E^5/E_{\rm UV}^2$ rather than with  $E^3$. Alternatively, the zero point part  $E^3$ is present, but is cancelled by the contribution at the compositeness scale $M_{cb}\sim M_{cf}$.   This would mean that there is a natural sum rule which nullifies the contribution of a composite object  leaving  only contribution of microscopic degrees of freedom to the vacuum energy (unless the Higgs condensate is formed).

 \section{Vacuum spectral function in condensed matter systems} 
   \subsection{Relativistic fermions emerging near Fermi surface}

It is instructive to consider the spectrum of vacuum energy in such effective 
relativistic theories
which emerge in the known microscopic background. 
For example the relativistic 1+1 dimensional fermions emerging near the Fermi surface
of interacting non-relativistic bare fermions in 1+1 dimension.
For illustration, we consider the simplified case of a mean-field model
of Fermi gas with attractive interaction, which can exist 
as isolated system without environment. This model ignores possible instability of the 
Fermi gas towards the Cooper pairing and other subtle phenomena occurring in 1+1 
dimension, which also simplifies the consideration. 
We choose the non-relativistic 
fermionic energy spectrum $p^2/2m$, and the energy density as a function 
of number density $n$ (non-relativistic analog of $q$ in
\\cite{KlinkhamerVolovik2008a,KlinkhamerVolovik2008b,KlinkhamerVolovik2008c})
in the form:
\begin{equation}
\epsilon(n) =  \epsilon_0(n)+\frac{1}{2\pi}\int_{-p_F}^{p_F} dp~  \frac{p^2}{2m}  
=  \epsilon_0(n) +\frac{\pi^2}{6m}n^3 ~~,~~ n=\frac{1}{2\pi}\int_{-p_F}^{p_F} dp= \frac{p_F}{\pi} ~,
\label{FermiSurfaceEnergyGen}  
\end{equation}
where $\epsilon_0(n)$ is the dominating term in the vacuum energy, while the 
kinetic part is considered as small perturbation.
We need the system, which can live without the external environment, and thus
has zero vacuum pressure in equilibrium.  
Condition of stability of equilibrium vacuum state in the absence of external pressure reads
\cite{KlinkhamerVolovik2008a}:
\begin{equation}
P_{\rm vac}=-\tilde{\epsilon}_{\rm vac}=- \epsilon(n) +n\frac{d \epsilon}{dn}= 0~.
\label{n_equil}  
\end{equation}
The equilibrium chemical potential is
\begin{equation}
\mu=\frac{d \epsilon}{dn}    =   \frac{p_F^2}{2m}+\frac{d \epsilon_0}{dn}~.
\label{mu}  
\end{equation}
The proper thermodynamic potential which corresponds to cosmological constant is
\begin{equation}
\Lambda=\tilde{\epsilon}_{\rm vac}=-P_{\rm vac}=\epsilon(n) -\mu n
=  \frac{1}{2\pi}\int_{-p_F}^{p_F} dp~\frac{p^2-p_F^2}{2m}  +\epsilon_0(n) -n\frac{d\epsilon_0}{dn}.
\label{FSEnergySpectrumGen}  
\end{equation}
This vacuum energy is the sum of the interaction term 
and energies of individual particles. Only the latter are relevant for the spectrum of vacuum energy. 
Let us now rewrite this in terms of the emergent relativistic fermions.
Relativistic fermionic quasiparticles  live near two Fermi surfaces
(Fermi points) at $p=\pm p_F$. Close to Fermi surface at $p=p_F$, the energy spectrum
of fermions is linear and relativistic:
\begin{equation}
 E=c\tilde p 
 ~,
\label{RelSpectrum}  
\end{equation}
where $c=v_F$ and $\tilde p=p- p_F$.
 In terms of the momentum $\tilde p$ counted from the Fermi surface, equation 
(\ref{FSEnergySpectrumGen})  reads
\begin{equation}
\Lambda=\Lambda_0+
  \frac{N_F}{2\pi}\int_{0}^{p_F} d \tilde p  \left(-v_F\tilde p + \frac{\tilde p^2}{2m} \right) =
\Lambda_0+ \frac{c}{\pi}\int_{0}^{p_F} d \tilde p  \left(- \tilde p + \frac{\tilde p^2}{2p_F}\right)  
 ~,
\label{FermiSurfaceEnergySpectrum2}  
\end{equation}
where $N_F=2$ is the number of emergent relativistic massless fermionic species,
emerging in the vicinity of the Fermi surface. The parameter $\Lambda_0$ plays the role of the bare cosmological constant: 
\begin{equation}
 \Lambda_0=\epsilon_0(n) -n\frac{d\epsilon_0}{dn} ~.
\label{BareCC}  
\end{equation}
This contribution to vacuum energy solely comes from
the microscopic physics of deep vacuum, and thus its energy spectrum does not make much sense.
Introducing the Planck energy  $E_{\rm P}=cp_F$ in equation 
(\ref{FermiSurfaceEnergySpectrum2}), one obtains
the following spectrum of $\Lambda$:
\begin{equation}
\lambda(E) =
 \frac{1}{\pi c} \left(- E + \frac{E^2}{2E_{\rm P}} \right) \Theta(E_{\rm P}-E)~~,~~
\Lambda=\Lambda_0+ \int_0^\infty dE~\lambda(E)=0~.
\label{FermiSurfLambdaSpectrum}  
\end{equation}
The first term in brackets in (\ref{FermiSurfaceEnergySpectrum2})  is the contribution of the Dirac sea
of two species of relativistic fermions with the spectrum $E=c\tilde p$. At low $E$ this is the leading term in the spectrum. After integration over all energy scales this  terms gives the traditional estimate 
for the Dirac sea energy in 1+1 dimesnions: 
\begin{equation}
  \epsilon_{\rm Dirac~vacuum}
=-2\int_0^{E_{\rm P}}\frac{dp}{2\pi} E= -\frac{E_{\rm P}^2}{2\pi c} ~.
\label{DiracVacuum}  
\end{equation}
The second term  in brackets in (\ref{FermiSurfLambdaSpectrum})
comes from the non-linear correction to the  Lorentz invariant spectrum (\ref{RelSpectrum}). At low energy $E\ll E_{\rm P}$ it is the subleading term in the 
spectrum, but at high energy this term gives the same order of magnitude as the Dirac sea,
$\epsilon_{\rm non-linear}=   E_{\rm P}^2/6\pi c$. In equilibrium vacuum both these  contributions
are compensated by the term $\Lambda_0$. 

\begin{figure}
\includegraphics[width=1.0\textwidth]{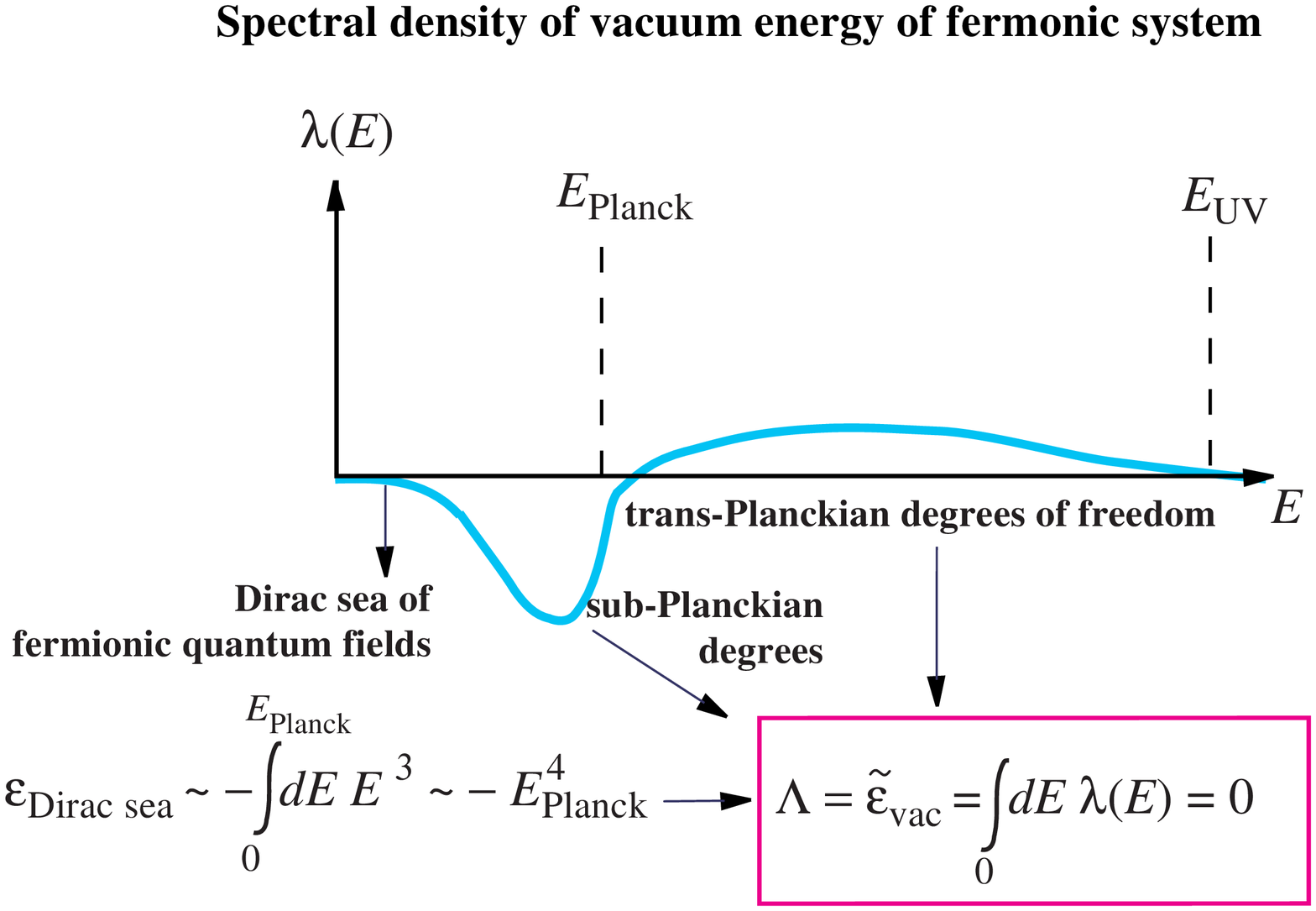}
\caption{Sectral density of $\Lambda$ suggested by
fermionic condensed matter. The Dirac sea energy of relativistic fermions
dominates at $E\ll E_{\rm P}$. It 
 diverges  as $-E^4$  producing contribution  $\sim -  E_{\rm P}^4$ to the vacuum energy. The contribution of the sub-Planckain region is fully compensated by a tiny response of the deep vacuum, whose characteristic energy scale $E \sim  E_{\rm UV}\gg E_{\rm P}$.}
\label{LambdaSpectrumF}
\end{figure}

The bare cosmological constant can be considered as the vacuum energy density originating from the  deep vacuum. To justify that,
let us consider the situation, when the term $\epsilon_0(n)$ in  (\ref{FermiSurfaceEnergyGen}) 
has characteristic energy scale $E_{\rm UV}$ much larger than the Planck energy, i.e. 
$ |\epsilon_0(n)| \sim E_{\rm UV}^2/\hbar c\gg E_{\rm P}^2/\hbar c$. In this case, the kinetic part
of $\epsilon(n)$ -- the second term in  (\ref{FermiSurfaceEnergyGen}) --  can be  considered as small perturbation.
Let us find the response of the deep vacuum to this perturbation which leads to the full compensation
of vacuum energy.
Let $n_0$ and $\mu_0$ be the equilibrium values of $n$ and $\mu$ of the quantum vacuum in zero approximation, when the kinetic term is absent. The obey  equations
\begin{equation}
-P_{\rm vac}=\epsilon(n_0) -\mu_0n_0=0~~,~~ \frac{d \epsilon}{dn}\Big|_{n=n_0}=\mu_0  
~.
\label{Equilibrium0}  
\end{equation}
Then from (\ref{n_equil})  it follows that in the first order approximation one has the following correction to the equilibrium density:
\begin{equation}
n=n_0(1+\alpha)~~,~~\alpha=-\frac{\pi^2}{3m}n_0^3 \chi=-\frac{E_{\rm P}^2}{3\pi c} \chi~.
\label{correction_to_n_0}  
\end{equation}
where $\chi$ is compressibility of the vacuum state \cite{KlinkhamerVolovik2008a,KlinkhamerVolovik2008b,KlinkhamerVolovik2008c}
\begin{equation}
\chi^{-1}=n_0^2 \frac{d^2\epsilon}{dn_0^2}>0~,
\label{compressibility}  
\end{equation}
Since $ 1/\chi \sim E_{\rm UV}^2/c$, the perturbation of the deep vacuum is small:   $|\alpha|  \sim  E_{\rm P}^2/E_{\rm UV}^2\ll 1$. The ultraviolet vacuum is so strong that even a tiny response
of this vacuum fully compensates the contribution of  sub-Planckian region.
Adding these ultraviolet degrees of freedom to $\lambda(E)$ one obtains the spectrum of $\Lambda$  illustrated in Fig.~\ref{LambdaSpectrumF}.

   \subsection{Relativistic fermions emerging near Fermi point} 

In a similar manner one can calculate the spectrum $\lambda(E)$ coming from the 3+1 relativistic Weyl fermions emerging in condensed  matter systems with topologically protected Fermi point. In case of $p$-wave superfluid (see Chapter 7 in Ref.  \cite{Volovik2003}),  we present omitting the derivation the leading contributions to $\lambda(E)$ coming in the low-energy region in vicinity of Fermi points:  
\begin{equation}
\lambda(E)\approx-N_F  \frac{E^3}{2\pi^2}\left(1-\frac{E}{2E_{\rm P}}\right)~~,~~E\ll E_{\rm P}~.
\label{FermiPoint}  
\end{equation}
Here $N_F$ is number of emergent relativistic massless fermionic species (for a  $p$-wave superfluid it is $N_F=2$); and we used such parameters of the system that all three Planck energy scales in Eq.(7.31) of Ref. \cite{Volovik2003} coincide. The leading $E^3$ term has the correct form, while the subleading term has power 4 instead of 5 in  (\ref{Micro}); it has the same relative magnitude as the
subleading term in (\ref{FermiSurfLambdaSpectrum}) and is normalized
to the Planck scale instead of the $E_{\rm UV}$ scale in (\ref{Micro}). This is because in both cases the underlying  vacuum -- the quantum liquid -- is non-relativistic even below the Planck scale, and becomes relativistic only in the vicinity of Fermi points, in the limit  $E\ll E_{\rm P}$. On the contrary,
our vacuum remains Lorentz invariant even above the Planck scale, i.e. even in the region 
$E_{\rm P}\ll E\ll E_{\rm UV}$
\cite{KlinkhamerVolovik2008a,KlinkhamerVolovik2008b}.

 \section{Discussion}

 The  consideration of condensed matter systems suggests that it is the trans-Planckian degrees of freedom of deep ultraviolet vacuum, which compensate the divergences of the vacuum energy occurring in the sub-Planckian
 region. In equilibrium, the back reaction of the deep vacuum leads to complete nullification of the vacuum energy responsible for the cosmological constant. This suggests that deep vacuum contributes also to the sum rules introduced by Zeldovich \cite{Zeldovich1968} for the spectrum of the vacuum energy, as we discussed in Sec. 
 \ref{Contribution_deep_vacuum}.  That is why the emergent theories do not support the idea  
 \cite{Kamenshchik2007,Alberghi2008} that
 the sum rules impose relations between the masses of the 
 fermionic and bosonic fields in Standard Model.
  
 The considered examples demonstrate a non-trival behavior of the spectrum of vacuum energy. 
 Massless relativistic fermionic quasiparticles  in condensed matter systems and graviton in the Frolov-Fursaev scheme of induced gravity both emerge form the constituent microscopic fields. However, we found that in the fermionic systems, the contribution of the emergent fermions to $\Lambda$ has a traditional form 
 at low energy, corresponding to the contribution of Dirac sea, Eqs.(\ref{FermiPoint}) and 
(\ref{FermiSurfLambdaSpectrum}). On the other hand,  the  Frolov-Fursaev scheme suggests that induced gravitons  do not contribute its zero point energy to $\Lambda$. This contradiction requires the further investigation of the spectrum of vacuum energy, in particular the contribution to the vacuum energy from the composite bosons and fermions must be studied.

I thank  Dmitri Fursaev,  Alexander Kamenshchik and Andrei Zelnikov for discussion.

\end{document}